# White matter hyperintensities volume and cognition: Assessment of a deep learning-based lesion detection and quantification algorithm on the Alzheimer's Disease Neuroimaging Initiative

Lavanya Umapathy[1,2], Gloria Guzman Perez-Carillo[2], Blair Winegar[3], Srinivasan Vedantham[4], Maria I. Altbach[4], and Ali Bilgin[1,4,5]

[1]Department of Electrical and Computer Engineering, University of Arizona, Tucson, Arizona, [2]Mallinckrodt Institute of Radiology, Washington University, St. Louis, MO, [3]Department of Radiology and Imaging Sciences, University of Utah, Salt Lake City, Utah, [4]Department of Medical Imaging, University of Arizona, Tucson, Arizona, [5]Department of Biomedical Engineering, University of Arizona, Tucson, Arizona

**ABSTRACT:** The relationship between cognition and white matter hyperintensities (WMH) volumes often depends on the accuracy of the lesion segmentation algorithm used. As such, accurate detection and quantification of WMH is of great interest. Here, we use a deep learning-based WMH segmentation algorithm, StackGen-Net, to detect and quantify WMH on 3D FLAIR volumes from ADNI. We used a subset of subjects (n=20) and obtained manual WMH segmentations by an experienced neuro-radiologist to demonstrate the accuracy of our algorithm. On a larger cohort of subjects (n=290), we observed that larger WMH volumes correlated with worse performance on executive function (P=.004), memory (P=.01), and language (P=.005).

**Keywords:** Convolutional Neural Networks, White Matter Hyperintensities, Cognition, ADNI

## INTRODUCTION:

White matter hyperintensities (WMH) are brain white matter lesions that appear bright in Fluid Attenuated Inversion Recovery (FLAIR) MR images[1]. The extent of WMH lesion burden is associated with degeneration of axons and myelin and is of clinical relevance in aging and age-related neurological disorders[2,3]. Increased WMH burden has been associated with a decline in cognitive factors such as executive function, memory, and language. Accurate detection and quantification of WMH and studying the temporal correlation between lesion burden and disease progression is of great interest in the neuroimaging community.

In this work, we use StackGen-Net[4,5], a fast and automated deep learning-based WMH segmentation algorithm, to detect and quantify WMH on isotropic resolution 3D FLAIR images. We use 3D FLAIR images from the Alzheimer's Disease Neuroimaging Initiative (ADNI) repository to demonstrate the accuracy of our segmentation algorithm (n=20). We also evaluate the clinical value of our algorithm by assessing the impact of WMH volumes on executive function, memory, and language on a group of ADNI subjects (n=290) diagnosed as cognitively normal (CN), mild cognitive impairment (MCI), and Alzheimer's Disease (AD).

## METHODS:

We recently proposed StackGen-Net, a stacked generalization ensemble of 3D Convolutional Neural Networks (CNNs), that improved segmentation of WMH in 3D FLAIR volumes compared to some state-of-the-art CNN segmentation frameworks[4,5]. StackGen-Net framework (Figure 1) consists of three orthogonal 3D CNNs (DeepUNET3D) each trained on axial, sagittal, and coronal reformatting of the 3D FLAIRs. Stacked generalization[6] maximizes the overall generalization accuracy by deducing the bias rate of individual CNNs in the ensemble. The Meta-CNN learns a new functional mapping from the WMH predictions of the individual orthogonal CNNs to the final WMH prediction. Trained on a cohort of subjects with a history of vascular disease, StackGen-Net segments WMHs in a time-efficient manner with performance comparable to human inter-observer variability[5].

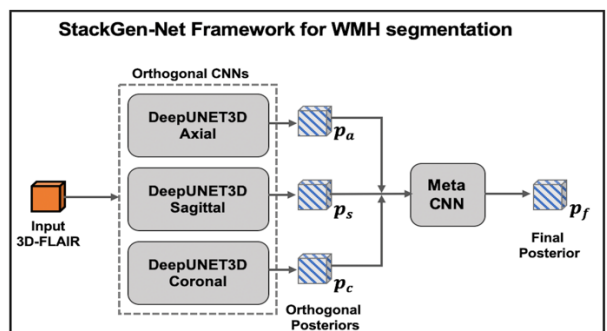

*Figure 1: Framework for StackGen-Net, a stacked generalization ensemble of Orthogonal Convolutional Neural Networks (CNNs). Each CNN predicted white matter hyperintensity (WMH) posterior probabilities on multi-planar orientations (axial, sagittal, and coronal) of a 3D FLAIR volume. The final WMH posterior is obtained using a Meta CNN that learns to combine individual CNN posteriors.*

Baseline data from subjects imaged with the ADNI-3 protocol (with 1mm isotropic 3D FLAIR volumes) were downloaded from ADNI website (www.adni.loni.usc.edu). This included demographics, diagnosis, education, WMH volume estimates, as well as composite scores for executive function (ADNI-EF), memory (ADNI-MEM), and language (ADNI-LAN) from the neuropsychological test battery[7]. The WMH lesion volumes from ADNI (ADNI-WMH) were quantified using a histogram-based technique[8].

The 3D FLAIR volumes from 290 subjects (193 CN, 73 MCI, and 24 AD) were brain-extracted and bias-corrected, followed by WMH detection and volume estimation by StackGen-Net (StackGen-Net-WMH). The total prediction time was 40s for a pre-processed 3D volume. An experienced neuro-radiologist manually annotated WMH on 3D FLAIR volumes of 20 non-demented participants selected randomly from this ADNI cohort to evaluate segmentation performance of our algorithm.

We used Bland-Altman analysis to compare agreement in WMH volumes from manual annotations to StackGen-Net-WMH and ADNI-WMH. The following statistics were also calculated: $R^2$, Coefficient of Variation (CV), and repeatability coefficient (RPC). We also used two-sided paired t-tests to see if StackGen-Net-WMH and ADNI-WMH volume estimates differed significantly from ground truth WMH volumes.

The associations of WMH volumes with ADNI-EF, ADNI-MEM, and ADNI-LAN were explored using multiple-linear regression models after adjusting for age, intercranial volume, sex, education level, APOE4 allele genotype, and diagnosis.

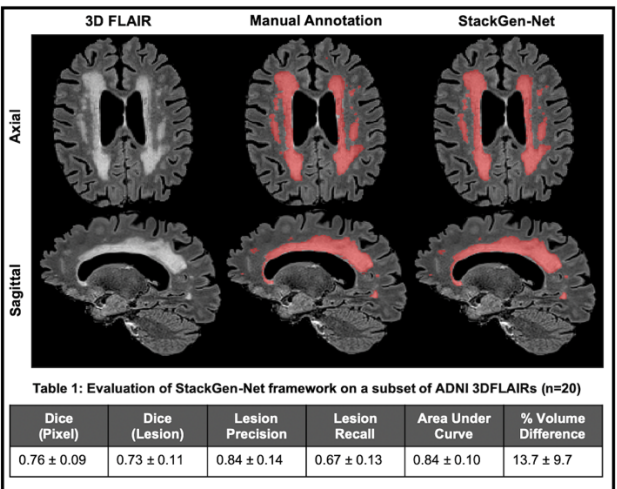

*Figure 2: Axial and coronal cross-sections of a test 3D-FLAIR volume from ADNI. The manual annotations are overlaid in the middle row for reference. The predictions from StackGen-Net agree well with manual annotations. Table 1 shows mean and standard deviation values for the different evaluation metrics on the ADNI test subset (n=20 volumes).*

## RESULTS:

Figure 2 shows WMH volume predictions on axial and sagittal cross-sections from a test ADNI 3D FLAIR volume. The WMH predictions from StackGen-Net agree well with manual annotations. Compared to manual annotations, StackGen-Net achieved average Dice score (pixel), Dice score (lesion), absolute volume difference, and area under precision-recall curve of 0.76 ± 0.09, 0.73 ± 0.11, 13.7% ± 9.7%, and 0.84 ± 0.10, respectively (n=20).

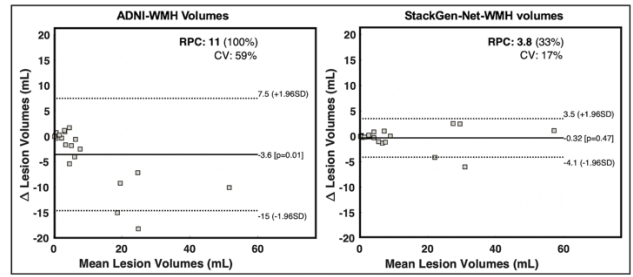

*Figure 3: Bland-Altman plots. The agreement between WMH volumes from manual annotations and two different segmentation algorithms is shown in these Bland Altman plots. Left: The WMH volumes were downloaded from ADNI and were estimated using an intensity histogram-based technique. Right: The WMH volumes were estimated by our deep learning-based framework. Notice that the limits of agreement are tighter in the plot on the right.*

Bland-Altman analysis (Figure 3) showed excellent agreement between ground-truth WMH volumes and StackGen-Net-WMH ($R^2$=0.98, CV=17%, RPC=33%). The limits of agreement for StackGen-Net-WMH were tighter when compared to ADNI-WMH ($R^2$=0.91, CV=59%, RPC=100%). We observed no significant WMH volume differences between StackGen-Net WMH volumes and ground truth (P=0.47, n=20, two-sided paired t-test). In contrast, this difference was significant (P=.01) for ADNI-WMH.

Multiple-linear regression models revealed that bigger volumes of WMH (StackGen-Net-WMH) were associated with worse performance on ADNI-EF (P=.004), ADNI-MEM (P=.01), and ADNI-LAN (P=.005). Similar, significant but less pronounced, effects were also observed using ADNI-WMH for ADNI-EF (P=.01), ADNI-MEM (P=.016), and ADNI-LAN (P=.016).

*Figure 2: ADNI cohort demographics (Table 2) and the P-values for the association between cognition and WMH volumes (Table 3) from the multiple linear regression models are shown.*

## DISCUSSION:

Many works using WMH volumes quantification on 2D-FLAIR images from ADNI have explored the relationship between WMH volumes and cognitive functions. The extent of the relationship (or lack thereof) often depends on the accuracy of the segmentation algorithm[9]. In this work, we used a subset of 3D FLAIR volumes (n=20) from ADNI and obtained manual segmentations of WMHs by an experienced neuro-radiologist to demonstrate the accuracy of our WMH segmentation algorithm. We also demonstrated, using a larger cohort of 3D FLAIR volumes (n=290) that larger WMH volumes correlate with significantly worse performance on executive function, memory, and language tasks, thereby affecting cognition. The analyses using WMH volumes from ADNI also agreed with the associations observed in this work.

## CONCLUSION:

The use of a stacked generalization of CNN models can provide a fast and accurate quantitative evaluation of WMHs to study association between WMH volumes and cognitive decline.

**Acknowledgements:** Arizona Health Sciences Center Translational Imaging Program Project Stimulus, BIO5 Team Scholar's Program, Arizona Alzheimer's Consortium. The data used in this work were obtained from Alzheimer's Disease Neuroimaging Initiative (ADNI) database (https://adni.loni.usc.edu).